# On single file and less dense processes

O. Flomenbom[*], A. Taloni[+,++]

[*] Department of Chemistry, Massachusetts Institute of Technology, Cambridge, MA, 02139
[+] Department of Physics, Massachusetts Institute of Technology, Cambridge, MA, 02139
[++] Institute of Physics, Academia Sinica - 128 Academia Road, Taipei, 11529, Taiwan.

**Abstract** - The diffusion process of $N$ hard rods in a 1D interval of length $L$ ($\rightarrow \infty$) is studied using scaling arguments and an asymptotic analysis of the exact $N$-particle probability density function (PDF). In the class of such systems, the universal scaling law of the tagged particle's mean absolute displacement reads, $<|r|> \sim <|r|>_{free}/n^{\mu}$, where $<|r|>_{free}$ is the result for a free particle in the studied system and $n$ is the number of particles in the covered length. The exponent $\mu$ is given by, $\mu = 1/(1+a)$, where $a$ is associated with the particles' density law of the system, $\rho \sim \rho_0 L^{-a}$, $0 \leq a \leq 1$. The scaling law for $<|r|>$ leads to, $<|r|> \sim \rho_0^{(a-1)/2}(<|r|>_{\text{free}})^{(1+a)/2}$, an equation that predicts a smooth interpolation between single file diffusion and free particle diffusion depending on the particles' density law, and holds for *any* underlying dynamics. In particular, $<r^2> \sim t^{\frac{1+a}{2}}$ for normal diffusion, with a Gaussian PDF in space for any value of $a$ (deduced by a complementary analysis), and, $<r^2> \sim t^{\frac{\beta(1+a)}{2}}$, for anomalous diffusion in which the system's particles all have the same power-law waiting time PDF for individual events, $\psi \sim t^{-1-\beta}$, $0 < \beta < 1$. Our analysis shows that the scaling $<r^2> \sim t^{1/2}$ in a 'standard' single file is a direct result of the fixed particles' density condition imposed on the system, $a=0$.

**Introduction.** - The *basic* single file problem is a diffusion problem of *N* hard rods (no bypassing is allowed) in an open 1D interval, namely, the system's length, *L*, goes to infinity, and the particles' density, $\rho$, is fixed, $\rho=\rho_0=N/L$, so $N=L/\Delta$, for some microscopic length scale $\Delta$ [1]. ($\Delta$ cannot be smaller than the particle's diameter, and can be taken as the average distance between the centers of nearest neighbor particles.) The underlying dynamics are homogenous, so for stochastic dynamics all the particles in the system have the same diffusion coefficient $D_0$. The statistics of a given particle, the tagged particle, in a basic stochastic single file, are known [2-9]. Hereafter, the tagged particle is identified by the coordinate *r*, and is taken to be the middle particle in the file (odd *N*). The probability density function (PDF) of the tagged particle is asymptotically Gaussian in position, with a variance that scales as the square root of time,

$$P(r,t|r_0) \sim (4Dt)^{-\frac{1}{4}} \exp\left\{-\frac{(r-r_0)^2}{\sqrt{4Dt}}\right\} \quad ; \quad D = D_0 \rho_0^{-2}. \quad (1)$$

The PDF in eq. (1) was first obtained in [2]. Subsequent studies provided complementary characterizations of the process [6-15], including the correlative motion of the particles in the file [15], the fluctuations in the particles' density [6, 15], the occurrence of a negative velocity auto-correlation function and non-pathological single event PDFs in time and space [14], and the system's normal behavior in higher dimensions [6]. For a deterministic single file with momentum exchange upon collisions, the tagged particle's PDF is also Gaussian but with a variance that scales as the time [3]. The single file process has also been associated with monomer dynamics in a large linear polymer: both systems share a similar scaling law for the mean square displacement of a tagged particle [23, 28].

The single file problem has attracted a lot of attention in recent years because it can model many real-world microscopic processes that now can be measured [24-32]. Examples include diffusion within biological and synthetic pores, and in porous materials, of water, ions, proteins, and organic molecules [24, 32]. Diffusion along 1D objects, such as the motion of motor-proteins along filaments [32]. Conductance in nano-wires [31]. However, in real-world systems, one, or several, of the conditions that define the *basic* single file process can break down. For example, real-world systems are finite, which means that a steady state regime is sure to be observed [20]. In an inhomogeneous system, the particles' diffusion coefficients are distributed with some PDF [21]. In a quasi-1D system e.g. a nano-channel, the particles may bypass each other with some constant probability upon collisions [16-19], and also in realistic channels, the particles can interact with the channel, meaning imposing (for example) a periodic potential on the file's particles [12].

Here, we relax the condition of the fixed particles' density, and use instead the scaling law, $\rho \sim L^{-a}$, $0 \leq a \leq 1$, which interpolates smoothly between the standard single file law ($a=0$) and a free diffusion ($a=1$). We find a universal scaling law for the tagged particle's mean absolute displacement, $<|r|> \sim <|r|>_{free}/n^\mu$, where $<|r|>_{free}$ is the result for a free particle in the studied system, $n$ is the number of particles in the covered length, and $\mu = \frac{1}{1+a}$. This scaling law describes systems with any underlying dynamics [e.g. normal diffusion, deterministic dynamics (super-diffusion), continuous time random walk dynamics (sub-diffusion)]. The scaling law for $<|r|>$ leads to, $<|r|> \sim \rho_0^{(a-1)/2}(<|r|>_{\text{free}})^{(1+a)/2}$, an equation that predicts a smooth interpolation between single file diffusion and free particle diffusion depending on the particles' density law, and holds for *any* underlying dynamics. In particular, $<r^2> \sim t^{(1+a)/2}$, for stochastic normal diffusion with a Gaussian PDF in space for any value of *a* (found in a complementary analysis), and, $<r^2> \sim t^{\beta(1+a)/2}$, for anomalous diffusive hard rods all having the same waiting time (WT) PDF for individual events, $\psi \sim t^{-1-\beta}$, $0 < \beta < 1$. Our analysis shows that the scaling $<r^2> \sim t^{1/2}$ in a standard single file results from the constant density condition imposed on the file (*a=0*).

**<|r|> for the tagged particle in a single file.** – It is well known that for a Brownian particle (i.e. a single particle random walk in an unconfined geometry and no traps), the mean absolute value scales as, $<|r|>_{free} \sim \sqrt{D_0 t}$, e.g. [33]. (Along the Letter, $\frac{<|r|>^2}{<r^2>} \to o(1)$ for large times.) For a tagged hard rod in the presence of hard rods in 1D, a slower expansion is expected because for a net distance to be covered by the tagged particle, the file's particles (in the relevant direction) must 'cooperate' and move in the direction of the propagation. Namely, the tagged particle evolution is a result of a correlative motion of the system's particles. These two basic properties are used below to derive a scaling law for $<|r|>$ for a tagged particle in a general class of processes of hard rods in 1D. We start by choosing $<|r|>_{free}$ as the natural length scale in the system. (This comes out in the rigorous calculations in the next subsection, but it is the only possible choice to begin with.) Then, the scaling argument must be proportional to one over the number of particles in the covered interval, *n*, raised to yet unknown power $\mu$ (assuming a non-decreasing scaling function). The more particles there are in the covered length, the harder is to achieve cooperation from them all. Thus, we write,

$$<|r|> = f\left(\frac{<|r|>_{free}}{n^{\mu}}\right).$$

$<|r|>_{free}$ is the upper bound on the diffusion for the tagged particle, so for $n=1$, $<|r|> = <|r|>_{free}$. Thus, we need to take a linear scaling function,

$$<|r|> \sim \frac{1}{n^\mu} <|r|>_{free}. \tag{2}$$

Note that $n$ in eq. (2) is a function of $<|r|>$ and not of $<|r|>_{free}$ because the number of particles in the actual covered distance is counted. Now, for the standard single file process, the particles' density is fixed, $n=\rho_0 r$, so, $<|r|> \sim \frac{1}{(\rho_0 <|r|>)^\mu}(D_0 t)^{1/2}$, and consequently,

$$<|r|> \sim \rho_0^{-\mu/(1+\mu)}(D_0 t)^{1/[2(1+\mu)]}. \tag{3}$$

Equation (3) leads to the known scaling law for standard single file when $\mu = 1$, implying,

$$<|r|> \sim \sqrt{\rho_0^{-1} <|r|>_{free}}, \tag{4}$$

valid for a standard single file system (constant density). Similar relation was obtained in ref. [10]. However, $\mu$ in eq. (2) can be a function of the particles' density, and this is shown in the next section.

Equations (2)-(4) imply the following: as eq. (4) is equivalent to, $<r^2> \sim \left(\frac{D_0}{\rho_0^2} t\right)^{1/2}$, the generalized diffusion coefficient reads, $D = D_0 \rho_0^{-2}$, and this result for $D$ coincides with $D$ in eq. (1). Also, the result, $<|r|> \sim \rho_0^{-1/2}(D_0 t)^{1/4}$, for a tagged particle in a standard single file is associated with three factors: (1) the requirement that the free particle mean absolute displacement scales as, $<|r|>_{free} \sim (D_0 t)^{\frac{1}{2}}$, (2) the requirement for 'cooperation' from the other particles in the length $<|r|>$ that scales as one over the number of particles in the covered length, and (3) the condition of a constant particles' density.

The scaling law in eq. (4) enables finding solutions for the mean square displacement for a tagged particle in related systems. For example, for the deterministic standard single file in which, $<|r|>_{\text{free}} \sim |v|t$, for some initial velocity $v$, $<|r|> \sim \sqrt{\rho_0^{-1}|v|t}$. Equivalent result was obtained in [3]. Another example is a system in which any particle performs a random walk with a WT-PDF that is not exponential, but decays as a power law, $\psi(t) \sim t^{-1-\beta}$, $0 < \beta < 1$. This can be the outcome of sticky walls or a branching geometry, and it is called here a CTRW (continuous time random walk) dynamics or just CTRW. (Note that the particles in the standard single file process have a single exponential WT-PDF for the individual transitions, i.e. $\psi(t) \sim e^{-t/\tau}$, for some microscopic timescale $\tau$.) For the CTRW model, $<|r|>_{\text{free}} \sim (\widetilde{D}_0 t)^{\frac{\beta}{2}}$ [34]. Using this relation in eq. (4) gives,

$$<|r|> \sim \rho_0^{-1/2} (\widetilde{D}_0 t)^{\frac{\beta}{4}},$$

as the density is still fixed. This result means that although all the particles move slowly, the single file nature of the process is still important: the tagged particle motion is still hindered by the presence of the file's particles, and cooperation from the file's particle is still required for a net evolution in a given direction to be obtained. Similar result for a CTRW dynamics with constant particles density was recently obtained in ref. [37]. Equation (2) can be a starting point in analyzing the effect of a general particles' density on the dynamics of the tagged particle. Consider a particles' density that obeys,

$$\rho = \frac{1}{\Delta}(L/\Delta)^{-a}, \quad 0 \leq a \leq 1, \tag{5}$$

where the limits on $a$ are expected: $a$ cannot be negative as the number of (classical) finite-size particles cannot be larger than the length of the system over the microscopic length Δ (in the same basic units); *a* is bounded by unity from above as *a>1* means a depletion of particles as the system's length increases, and such a process is not considered here. The value *a=0* corresponds to the standard single file process, and the value *a=1* corresponds to the free diffusion process. Strictly speaking, $\rho$ in eq. (5) is the initial density of the file: the particles are positioned at, $x_{0,j} = sign(j)\Delta |j|^{1/(1-a)}$, for $|j| \leq L^*$, where $L^*$ is arbitrary but finite. For distances larger than $L^*$ the density is fixed, $\rho^* = \frac{1-a}{\Delta}(L^*/\Delta)^{-a}$. We define $\rho^*$ (and $L^*$) to get, for large times, a standard single file dynamics (as $L^* \ll L$), rather than a free particle behavior. For intermediate times (to be found below), diffusion faster than the standard single file regime but slower than free diffusion is expected, reflecting the expansion process from dense to dilute environments occurring around the tagged particle. To derive the scaling law for $<|r|>$ in this regime, we first translate the relation for $\rho(L)$ into an equation for the number of particles as a function of length, $n \sim (\rho_0 L)^{1-a}$. Then, we need to estimate the dependence of $\mu$ in eq. (2) with *a*. This is done in the next section by deriving the PDF for the tagged particle for normal diffusion in a system obeying the density law in eq. (5). This analysis finds $\mu(a)$ for any underlying dynamics because the underlying dynamics is not correlated with the density law.

**The PDF of the tagged particle.** – To derive the PDF for a tagged particle in a normal diffusion system, we start with the equation of motion for the joint *N*-particle PDF (no inertia),

$$\partial_t P(\{x\}, t|\{x_o\}) = D_o \sum_{j=-M}^{M} \partial_{x_j^2} P(\{x\}, t|\{x_o\}) , \tag{6}$$

where $2M+1=N$, and $\{x\}=x_{-M},...,x_M$. In the following, we take $x_{o,j} = \Delta j$ (unless otherwise is explicitly noted). The single file nature enters through the system's boundary conditions,

$$[\partial_{x_j}P(\{x\},t|\{x_o\}) - \partial_{x_{j+1}}P(\{x\},t|\{x_o\})]_{x_j=x_{j+1}} = 0 \quad ; \quad -M \leq j < M,$$

which simply means that the adjacent particles cannot bypass each other. The solution for eq. (6) can be obtained from the Bethe ansatz [35-36]. For this problem, the Bethe ansatz gives the $\{k\}$-space, $\{k\}=k_{-M},...,k_M$, integrand of the Fourier transform of the solution (x→k) [20],

$$\hat{P}(\{k\},t|\{x_o\}) = \frac{1}{N!}\sum_p e^{-\sum_{j=-M}^{M} ik_j(x_{0,j}-x_j(p))+D_0tk_j^2}.$$

Here, the index $p$ stands for the $p$ permutation of the $N$ particles' indices, so the summation is over $N!$ permutations (e.g. $x_j(p)=x_i$, for a given $p$, and, $-M \leq i,j \leq M$). The normalized joint PDF in $\{x\}$-space reads,

$$P(\{x\},t|\{x_o\}) = (4\pi D_0 t)^{-\frac{N}{2}} \sum_p \prod_j e^{\frac{-(x_j-x_{o,j}(p))^2}{4D_0 t}}.$$

To show that $P(\{x\},t|\{x_o\})$ is normalized to one, we need to perform an $N$-dimensional integration over the $\{x\}$-space with the restriction,

$$-\infty \leq x_{-M} \leq x_{-M+1} \leq \cdots \leq x_{M-1} \leq x_M \leq \infty. \tag{7}$$

It is easily seen from the direct calculations for small $N$ values that the restricted integration can be replaced by an unrestricted integration for each particle, i.e., $-\infty \leq x_j \leq \infty$, $j=-M,...,M$, when dividing by $N!$. Each permutation in the expression for $P(\{x\},t|\{x_o\})$ is a product of $N$ integrals, each of which is normalized to one. Thus, each permutation contributes a factor of $1/N!$. As there are $N!$ permutations, $P(\{x\},t|\{x_o\})$ is normalized to one.

To obtain the PDF for the tagged particle, $P(r,t|r_0)$, $r \equiv x_0$ and $r_0 = 0$, we need to integrate out all the file particles' coordinates except for $r$, while obeying the restrictions of eq. (7). This is done when separating the integrals into left integrals and right integrals (relative to tagged particle),

$P(r,t|r_0) =$

$\int_{-\infty}^{x_{-M+1}} dx_{-M} \int_{-\infty}^{x_{-M+2}} dx_{-M+1} \ldots \int_{-\infty}^{r} dx_{-1} \int_{r}^{\infty} dx_1 \int_{x_1}^{\infty} dx_2 \ldots \int_{x_{M-1}}^{\infty} P(\{x\},t|\{x_o\}) dx_M$.

This *2M*-dimensional integration fulfills eq. (7); the particles always keep their relative position. Similar to the manipulation done in the calculations of the normalization constant, we can change all the upper bounds in the left integrals to $r$, and also change all the lower bounds in the right integrals to $r$. Thus, $P(r,t|r_0)$ is given by,

$$P(r,t|r_0) = \frac{1}{C} \prod_{j=1}^{M} \int_{-\infty}^{r} dx_{-j} \int_{r}^{\infty} dx_j \, P(\{x\},t|\{x_o\}), \tag{8}$$

where $C$ is the normalization constant. Equation (8) enables further analysis because it gives $P(r,t|r_0)$ as products of separate integrals,

$$P(r,t|r_0) \propto \sum_p e^{-\frac{1}{r_f^2}[r-r_0(p)]^2} \prod_{j=1}^{M} \int_{-\infty}^{r} dx_{-j} \, e^{-\frac{1}{r_f^2}[x_{-j}-x_{0,-j}(p)]^2} \int_{r}^{\infty} dx_j \, e^{-\frac{1}{r_f^2}[x_j-x_{0,j}(p)]^2}.$$

Here, for notation convenience, we define, $r_f \equiv \sqrt{4D_0 t}$. ($r_f$ equals $<|r|>_{free}$ for normal diffusion. As stems from the calculations below, $r_f$ is the natural length scale in the system.) For any permutation, the faith of each integral over $x_j$, with $j>0$, is one of three possible outcomes (asymptotic analysis):

(1) When $(r - x_{0,j})/r_f \to 0$, the integral is approximated by, $\sqrt{\pi}/2$.

(2) When $(r - x_{0,j})/r_f \to -\infty$, the integral is approximated by, $\sqrt{\pi}$.

(3) When $(r - x_{0,j})/r_f \to +\infty$, the integral is approximated by, $\frac{e^{-Y_j^2}}{2|Y_j|}$, where, $Y_j = (r - x_{0,j})/r_f$.

The same three possible outcomes are obtained for any integral over $x_j$ with $j<0$, when switching the condition-part of cases (2) and (3) above. For each permutation, we need to count the number of integrals of each kind, and then to sum over all permutations' results. We continue by firstly analyze $P(r,t|r_0)$ for small values of $r$. Here, small $r$ values means, $|r| \leq r_f$. We define ordered permutations as permutations in which all the positive initial conditions are to the right of $r$ and all the negative initial conditions are to the left of $r$. For small $r$ values, there are $(M!)^2$ such permutations: there are $M!$ internal permutations of the left initial conditions and $M!$ internal permutations of the right initial conditions, starting from the 'perfectly' ordered permutation, $p=1$: $x_{0,j}(1) = \Delta j$ for every $j$. All $(M!)^2$ permutations of the 'perfectly' ordered permutation lead to the same result, as the integrals in eq. (8) are separated. For small $r$, only cases (1) and (2) are relevant for the ordered permutations, so each ordered permutation gives a constant independent of $r$. Note that the actual value of $r_0(p)$ is irrelevant: all the 'ordered-permutations' contribute a constant, and the summation over $p$ can be replaced by the integral, $\sum_p e^{-\frac{1}{r_f^2}[r-r_0(p)]^2} \sim \int_{-\infty}^{\infty} e^{-\frac{1}{r_f^2}[r-\Delta p]^2} dp$, which is independent of $r$. This is a general result not limited to the ordered permutations, or small $r$ values: the effect of an individual initial condition on the result can be safely neglected in the thermodynamics limit of $N \to \infty$ particles. The initial condition for the tagged particle enters while counting type (3) integrals for 'not-ordered' permutations. Similar effect is shown below in the calculations for large $|r|$.

We calculate now the $4^M (M!)^2$ permutations in which the initial conditions are not ordered. We choose *m* initial coordinates from the left *M* initial ordered coordinates and *m* initial coordinates from the right *M* ordered initial coordinates, and switch between the two sets. For each switch, there are the 'standard' $(M!)^2$ internal permutations all lead to the same result (to be calculated for each switching protocol). We distinguish between two choice types: the chosen coordinate is within the distance $r_f$ from *r* or not. So, there are 4 possibilities for each *switch*. The contribution from switching an initial coordinate within the distance of $r_f$ from *r* with an initial coordinate within the distance of $r_f$ from *r* from the other side gives approximately the result of the ordered permutations discussed above, i.e. the result is independent of *r*. The contribution from permutations in which both initial coordinates that are switched are more distant than $r_f$ from *r* (in opposite direction) vanishes. The important case is when an initial coordinate within the distance of $r_f$ from *r* is switched with an initial coordinate from the other side (right-left switch or left-right switch) that its distance to *r* is larger than $r_f$. For such 'mixed' cases, the overall contribution is proportional to,

$$P(r,t|r_0) \propto (M!)^2 \sum_{z=1}^{\rho_0 r_f} \sum_{q=1}^{\rho_0 r_f - z} S_z S_q \prod_{j=1}^{z} \frac{e^{-[Y_j(p)]^2}}{|Y_j(p)|} \prod_{i=1}^{q} \frac{e^{-[Y_{-i}(p)]^2}}{|Y_{-i}(p)|}, \tag{9}$$

In the upper bounds of the summations in eq. (9), we translated distances into particle numbers using the density. Also, we define in eq. (9) the combinatorial factor, $S_z = \binom{M - \rho_0 r_f}{z} \binom{\rho_0 r_f}{z}$, which gives the number of ways to perform the switching protocol for *z* coordinates. Equation (9) has two combinatorial factors, one for switching right initial coordinates within the distance $r_f$ from *r* with distant left initial coordinates, and the second for switching left initial coordinates within the distance $r_f$ from *r* with distant right initial coordinates. Each combinatorial factor is associated with a product of Gaussians resulting from the integrations of case (3). In the thermodynamic limit, $M$ is much larger then $r_f$, and the symmetric term, $z = q = \frac{\rho_0}{2} r_f = \rho_0 \sqrt{D_0 t}$ dominates the sum in eq. (9). So we have,

$$P(r,t|r_0) \propto \left(M! \, S_{\rho_0 \sqrt{D_0 t}}\right)^2 \prod_{j=1}^{\rho_0 \sqrt{D_0 t}} \frac{e^{-[Y_j(p)]^2 - [Y_{-j}(p)]^2}}{|Y_j(p)||Y_{-j}(p)|}$$

$$\lesssim \left(M! \, S_{\rho_0 \sqrt{D_0 t}}\right)^2 e^{-\rho_0 \sqrt{D_0 t}[Y_+^2 + Y_-^2 + \log|Y_-||Y_+|]}$$

where $Y_\pm = \frac{r \pm r_f}{r_f}$. In the second line, we approximate the Gamma function by the Stirling formula, and used the fact that the all initial coordinates in the $Y_j$s are more distant than $r_f$ from $r$. Thus, the leading term for the PDF of the tagged particle is given by,

$$P(r,t|r_0) \propto e^{-\frac{r^2}{\sqrt{4Dt}}}, \tag{10}$$

with a logarithmic correction in the exponent. For large values of $r$, $|r| \geq r_f$, there are always $\rho_0 r$ initial coordinate to the left of $r$ (say $r>0$). This gives rise to a correction term, $e^{-\frac{\rho_0(|r|-\sqrt{4D_0 t})^3}{4D_0 t}}$, which is multiplied by the result of any permutation. The switching analysis, however, is the same as discussed above. Thus, the tagged particle's PDF for $|r| \geq r_f$ reads,

$$P(r,t|r_0) \propto e^{-\frac{r^2}{\sqrt{4Dt}} - \frac{\rho_0(|r|-\sqrt{4D_0 t})^3}{4D_0 t}}.$$

The correction term is important only when $|r| \geq 3r_f$, but the PDF at such distances is of the order of $O(10^{-6})$.

To obtain the tagged particle PDF for less dense systems, we note that in the above calculations, the fixed particles' density condition entered in the upper bounds in the sums in eq. (8). For less dense systems, the upper bound in eq. (8), $\rho_0 r_f$, is replaced by $(\rho_0 r_f)^{(1-a)}$ because the basic length scale in the underlying Gaussians is still $r_f$. The result for less dense systems reads,

$$P(r,t|r_0) \propto e^{-\frac{(r/\rho_0^a)^2}{(4Dt)^{(1+a)/2}}}. \tag{11}$$

Equation (11) is the tagged particle's PDF (to a leading term) for any $r$ and any particles' density that obeys eq. (5).

**<|r|> for the tagged particle in less dense systems.-** The scaling law for $<r^2>$ calculated from eq. (11) reads, $\rho_0^2 <r^2> \sim (\rho_0^2 D_0 t)^{(1+a)/2}$. This relation can be written as a general relation,

$$<|r|> \sim \rho_0^{(a-1)/2} (<|r|>_{free})^{(1+a)/2}. \tag{12}$$

Equation (12) is equivalent of setting $\mu = \frac{1}{1+a}$ in eq. (2), resulting in,

$$<|r|> \sim \frac{1}{n^{1/(1+a)}} <|r|>_{free}. \tag{13}$$

Equation (13) emphasizes the fact that the particles in the covered length hinder the diffusion of the tagged particle less and less as the system itself becomes less dense.

Equation (12) explains the relation, $<|r|> \sim \sqrt{\rho_0^{-1} <|r|>_{free}}$, obtained for $a=0$: this relation is a direct result of the fixed density condition imposed in a standard single file. Equation (12) also generalizes it to any density law, and predicts a smooth interpolation between single file diffusion and free particle diffusion for any underlying dynamics Equation (12) enables a clear cut identification of the contributions to $<|r|>$ from the particles' density law and from the underlying dynamics. For the particular cases of normal and CTRW dynamics, we find from eq. (12),

$$< |r| > \sim \begin{cases} \rho_0^{-1}(\rho_0^2 D_0 t)^{\frac{1+a}{4}} & normal\ diffusion \\ \rho_0^{-1}\left(\rho_0^{2/\beta}\widetilde{D}_0 t\right)^{\frac{\beta(1+a)}{4}} & CTRW \end{cases}. \quad (14)$$

(For free diffusion, $\rho_0$ in eq. (14) is set to unity). These relationships are valid up to the exit time from the interval $L^*$, $\rho_0^2 D_0 t^* \sim c(\rho_0 L^*)^{4/(1+a)}$, for normal diffusion, and $\rho_0^{2/\beta}\widetilde{D}_0 t^* \sim c^{1/\beta}(\rho_0 L^*)^{4/[(1+a)\beta]}$, for CTRW, where $c = (1-a)^{2(1-a)/[(1+a)a]}$. For times larger than $t^*$, a standard single file scaling for $<|r|>$ is expected, with a constant density, $\rho^* = \frac{1-a}{\Delta}(L^*/\Delta)^{-a}$. For example, taking $\rho_0 L^* = 10^3$ and $a \approx 1/3$, the scaling law in eq. (14) spans three orders of magnitude in the dimensionless time parameter, $\rho_0^2 D_0 t$. Results from simulations for normal diffusive hard rods confirm eq. (12) (Fig. 1). Note that it stems from eq. (14) that a process with $<|r|> \sim t^{1/4}$ can be a consequence of three different scenarios: (1) a free particle ($\rho \sim L^{-1}$) CTRW process with $\psi \sim t^{-1-1/2}$, (2) a single file process ($\rho \sim \rho_0$) with an exponential microscopic WT-PDF, (3) a combination of a CTRW process ($\psi \sim t^{-1-\beta}$) and a density that decays with the distance ($\rho \sim L^{-a}$), such that $\beta = \frac{1}{1+a}$.

Lastly, note that when the tagged particle has a different underlying dynamics than all the other particles in the file, it adopts the dynamics of the file, eq.(12), unless its underlying dynamics is slower than eq.(12), and in this case, a free particle scaling law is observed. For example, consider a case where the tagged particle's underlying dynamics is a CTRW, $\psi(t) \sim t^{-1-\beta}$, and the single file particles have a single exponential WT-PDF. Only when, $\beta \leq \frac{1+a}{2}$, $<|r|> = <|r|>_{\text{free}}$. For, $\beta > \frac{1+a}{2}$, the tagged particle follows the single file evolution, and $<|r|>$ is given by the first line on the right hand side of eq. (14).

**Concluding remarks. -** This Letter deals with the dynamics of hard rods in an infinite 1D system. The first main result in this Letter is the general scaling law for the mean absolute displacement for a tagged particle, $<|r|> \sim <|r|>_{free}/n^{\mu}$. Here, $<|r|>_{free}$ is the result for a free particle in the studied system, $n$ is the number of particles in the covered length, and $\mu = \frac{1}{1+a}$ for a system with an initial density law, $\rho = \frac{1}{\Delta}(L/\Delta)^{-a}$, $0 \leq a \leq 1$. The factor $1/n^{\mu}$ is associated with the demand for cooperation from the hard rods in the covered length. The scaling law for $<|r|>$ enables deriving a general estimation for $<|r|>$ for the tagged particle for systems with any particles' density and any underlying particles' dynamics: $<|r|> \sim \rho_0^{(a-1)/2}(<|r|>_{free})^{(1+a)/2}$. This estimation is the second main result in this Letter. It shows that varying the particles' density law leads to a smooth interpolation between a 'standard' single file scaling and a free particle scaling for the mean absolute displacement. Specific results for normal diffusion and CTRW dynamics are given in eq. (14), where it was also shown that for normal diffusive hard rods, the tagged particle PDF is always a Gaussian (to a leading order), eq. (11), for any particles' density law that obeys eq. (5). Finally, it was shown that when the tagged particle diffuses according to a different underlying dynamics than all the other particles in the system, it adopts the dynamics of the other particles in almost all cases. Otherwise, a free diffusion scaling is observed.

***

We acknowledge discussions with R. Metzler, M. A. Lomholt, L. Lizana, F. Marchesoni and M. Kardar. This work was partially supported by the NSF under grant CHE 0556268. T.A. acknowledges the support from the Knut and Alice Wallenberg foundation.

**Figure captions**

Stochastic simulations of single file dynamics. A log-log plot of the mean square displacement of a tagged particle (middle particle) in a single file of $N$ particles as a function of time. The initial particles' density obeys eq. (5). The size of the system is a constant for all curves (4000 lattice spacing), and the actual value of $N$ depends on the initial particle's density. 1000 runs were performed for each value of the power $a$ that appears in the initial density law. The input value of $a$ is indicated for each curve together with the value of the fit-power denoted by $\gamma$. Convergence of $\gamma$ to the estimation $(1 + a)/2$ is observed for all the curves.

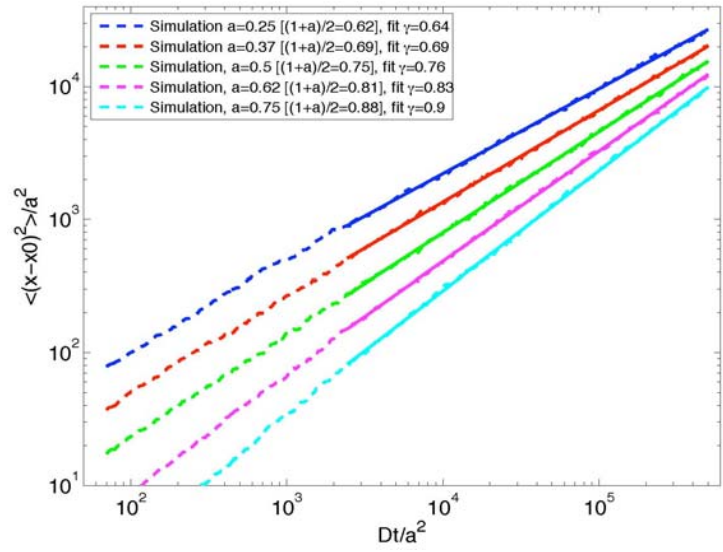